\documentclass[fleqn,usenatbib,useAMS]{mnras}

\usepackage{graphicx}	
\usepackage{amsmath}	
\usepackage{amssymb}	
\usepackage{multicol}        
\usepackage{bm}		
\usepackage{pdflscape}	

\usepackage{multirow}

\usepackage[T1]{fontenc}
\usepackage{ae,aecompl}

\usepackage{newtxtext,newtxmath}

\title[Monoperiodicity of BLAPs]{Confirmation of Monoperiodicity Above $20$ Seconds for\\ Two Blue Large-Amplitude Pulsators}

\author[P. R. McWhirter et al.]{Paul Ross McWhirter$^1$\thanks{Contact e-mail: \href{mailto:P.R.McWhirter@ljmu.ac.uk}{P.R.McWhirter@ljmu.ac.uk}},
Marco C. Lam$^{1,2}$ and
Iain A. Steele$^1$
\\
$^1$Astrophysics Research Institute, Liverpool John Moores University, IC2, LSP, 146 Brownlow Hill, Liverpool L3 5RF, U.K.\\
$^2$Astronomical Observatory, University of Warsaw, Al. Ujazdowskie 4, 00-478, Warszawa, Poland}

\date{Last updated \today}

\pubyear{2020}

\begin{document}
\label{firstpage}
\pagerange{\pageref{firstpage}--\pageref{lastpage}}
\maketitle

\begin{abstract}

Blue Large-Amplitude Pulsators~(BLAPs) are a new class of pulsating
variable star. They are located close to the hot subdwarf branch in
the Hertzsprung-Russell diagram and have spectral classes of late O
or early B. Stellar evolution models indicate that these stars are
likely radially pulsating, driven by iron group opacity in their
interiors. A number of variable stars with a similar driving
mechanism exist near the hot subdwarf branch with multi-periodic
oscillations caused by either pressure (p) or gravity (g) modes.
No multi-periodic signals were detected in the OGLE discovery light
curves since it would be difficult to detect short period signals associated with higher-order p modes
with the OGLE cadence. Using the RISE instrument on the Liverpool
Telescope, we produced high cadence light curves of
two BLAPs, OGLE-BLAP-009 ($m_{\mathrm{v}}=15.65$\,mag) and
OGLE-BLAP-014~($m_{\mathrm{v}}=16.79$\,mag) using a $720$\,nm 
longpass filter. Frequency analysis of these light curves identify
a primary oscillation with a period of $31.935\pm0.0098$\,mins and an
amplitude from a Fourier series fit of $0.236$\,mag for BLAP-009.
The analysis of BLAP-014 identifies a period of $33.625\pm0.0214$\,mins and
an amplitude of $0.225$\,mag. Analysis of the residual light curves reveals no
additional short period variability down to an amplitude of $15.20\pm0.26$\,mmag
for BLAP-009 and $58.60\pm3.44$\,mmag for BLAP-014 for minimum periods of $20$\,s and
$60$\,s respectively. These results further confirm
that the BLAPs are monoperiodic.

\end{abstract}




\begingroup
\let\clearpage\relax
\endgroup
\newpage

\section{Introduction}
Stellar pulsation is a phenomenon witnessed across the
Hertzsprung-Russell~(H--R) diagram from the main
sequence~(MS) through to the white dwarf~(WD) sequence~\citep{1995ARA&A..33...75G, 1996ARA&A..34..551G}.
These pulsations occur when stars of given composition and
structure expand and radially contract their outer layers to maintain
equilibrium~\citep{1917Obs....40..290E} or due to non-radial variations in surface temperature~\citep{1989nos..book.....U}. They can be used to probe the inner structure of
these variable stars~\citep{1984ARA&A..22..593D}. Large
amplitude, radially pulsating stars include the common
$\delta$ Scuti A-type MS stars~\citep{1900ApJ....12..254C}, the metal-poor blue straggler SX Phoenicis stars~\citep{1952PASP...64...31E, 1955AJ.....60..179S}, the primarily G-type bright giant classical Cepheids~\citep{1786RSPT...76...48G} and the metal-poor A and
F-type giant RR Lyrae stars~\citep{1901ApJ....13..226P}.

In these variable stars, the strongly periodic radial pulsations
are due to the $\kappa$--mechanism, an opacity change caused by
ionization of He\,II within the stellar interior leading to a
build up of energy which produces a cycle of expansion and contraction~\citep{1917Obs....40..290E}.
Radial pulsations produce a clear signal in stellar light
curves in the shape of a periodic sawtooth or sinusoidal 
variation~\citep{1991PASP..103..933M}.

The light curve shape depends on the effective
temperature of the star and the filter band used for the measurement. 
Shorter wavelength bands are dominated by the variation of the star's effective temperature and have a
sawtooth shape whereas longer wavelengths are more sinusoidal and are primarily due to radial variations.
Wavelengths between these two ranges show a contribution from both processes. The definition of shorter
and longer wavelengths in this explanation is determined by the effective temperature of the star. For
hot pulsators, the temperature-dominated light curves are found in the ultraviolet bands, and the
radius-dominated light curves are in the optical bands. For cooler pulsators, these effects are found in
longer wavelength passbands with the location of temperature-dominated light curves in the optical bands
and radius-dominated light curves are in the near infrared.

Blue large-amplitude pulsators (BLAPs) are a new class of
variable star~\citep[hereafter P17]{2017NatAs...1E.166P} identified by the
Optical Gravitational Lensing Experiment~(OGLE) within fields pointing
to the Galactic Bulge~\citep{2008AcA....58...69U, 2015AcA....65....1U}. In P17, the
prototype of the class was initially identified as a
$\delta$ Scuti star due to its short period of
$20$-$40$\,mins, variability amplitudes of $0.19$-$0.36$\,mag
in the I-band and $0.22$-$0.43$\,mag in the V-band and
sawtooth-shaped light curves.
P17 shows there is a clear periodic colour change as a function of the oscillation phase 
folded at the dominant period suggesting
pulsation is the cause of the variability.
Their amplitudes are similar to the High Amplitude
$\delta$ Scuti~(HADS) variables but their periods differ as
HADS stars have longer periods ($1.2$--$4.8$\,hrs)~\citep{2006MmSAI..77..223P}. The spectroscopic
followup by P17 confirmed that BLAPs are substantially hotter
than $\delta$ Scuti variables with effective temperatures of
$\mathrm{T}_{\mathrm{eff}}\approx30,000$\,K, surface gravity of 
$\log g / ({\rm cm s^{-2}}) = 4.4 - 4.8$ and
moderate helium enrichment.
The OGLE survey also included multiple fields aimed at the
Magellanic Clouds and followup study of these fields has revealed no BLAPs~\citep{2018pas6.conf..258P}.
After the discovery of the BLAPs, a second similar type of
variable star was identified: high-gravity
BLAPs~\citep{2019ApJ...878L..35K}. These stars have a similar effective
temperature but higher surface gravity than the original BLAPs,
shorter periods of $200$-$500$\,s, amplitudes of
$0.1$-$0.2$\,mag in the ZTF-r-band~\citep{2019PASP..131a8002B} and more sinusoidal light curves.
Followup spectroscopic observations show a periodic colour change and radial velocity variation
as a function of oscillation phase indicating the variability is due to a pulsating atmosphere~\citep{2019ApJ...878L..35K}.

The stellar parameters reported by P17 place BLAPs at a sparsely populated location
bluer than the MS and
above the WD sequence on the H--R diagram. They are early B-type/late O-type stars close to the hot subdwarf (sdOB) branch but with surface gravity ten times lower, and higher luminosity, indicating that they are in a giant configuration.
Oscillation analysis of stellar evolution models of hot subdwarfs, 
with a similar temperature but higher surface gravity than BLAPs,
indicate there are pressure (p) and gravity (g) mode instabilities
which can drive non-radial pulsations~\citep{2006MNRAS.371..659J}.
These modes are responsible for the pulsating classes of hot subdwarf EC14026~($90-600$\,s periods)~\citep{1997MNRAS.285..640K} and PG1716~($45-180$\,min periods)~\citep{2003ApJ...583L..31G}. Other OB-type pulsating
stars include the MS $\beta$ Cephei B-type giant
stars~\citep{1902ApJ....15..340F} and the slowly
pulsating B~(SPB) stars~\citep{1985A&A...152....6W}.
These pulsations are also due to opacity changes in their stellar
atmospheres but instead of helium~\citep{1996ARA&A..34..551G},
it is due to the partial ionization of iron group atoms at
temperatures of $200,000$\,K in the interior of these
stars~\citep{1989BAAS...21.1095C, 1993MNRAS.262..204D, 1993MNRAS.265..588D, 2006MNRAS.371..659J}.

Stellar evolution models indicate that
BLAPs are likely either extremely low mass~(ELM) pre-WD stars~\citep{2018MNRAS.477L..30R}
or higher mass core helium-burning stars evolving to the hot
subdwarf branch~\citep{2018MNRAS.478.3871W}. In the case of a pre-WD stellar evolution stage, the evolution of an $\approx1\mathcal{M}_{\sun}$ mass zero age main sequence (ZAMS) star of solar metallicity or greater can lead to a $0.27$-$0.37\mathcal{M}_{\sun}$ pre-ELM WD~\citep{2018MNRAS.477L..30R}. Such a star would exhibit effective temperature and surface gravity similar to those observed in BLAPs with extended envelopes and hydrogen-burning shells. Mixing due to some combination of convection, rotation and shell-flashes results in the stellar surface containing a mixture hydrogen and helium. The pre-ELM WD models indicate that observed pulsation periods can be produced by high-order non-radial g modes at longer periods and the fundamental radial modes at short periods~\citep{2019A&ARv..27....7C}. \citet[hereafter BJ18]{2018MNRAS.481.3810B} included the atom diffusion process of iron group elements in their stellar evolution models computed with Modules for Experiments in Stellar Astrophysics~(\texttt{MESA})
software~\citep{2010ascl.soft10083P}. This leads to the radiative levitation of those heavy chemical species, due to differential forces on atoms with different opacity. Their simulation shows an enriched iron group layer in the interior of low-mass pre-WDs that can drive a fundamental mode pulsation across the observed BLAP period range.
The rate of period change poses a problem for the fundamental radial mode as it
should always be negative for increasing stellar age which disagrees
with the positive and negative values
observed by P17. Non-radial g mode
pulsations can exhibit both positive and negative rates of period
change more similar to the observations~\citep{2019A&ARv..27....7C}.

P17 also considers the alternative model of a core helium-burning pre-sdOB star requires
a higher mass progenitor of
$5\mathcal{M}_{\sun}$. Whilst less MS stars evolve with
this required mass, the stellar evolution models show they do cross the BLAP instability
region. More importantly, the observed values of the rate of period change for the BLAPs
are $10^{-7}-10^{-8}$\,$\mathrm{yr}^{-1}$ which agrees with the
fundamental mode of the core helium-burning
stars compared to the pre-ELM WD models which have a rate of period change of
$10^{-5}$\,$\mathrm{yr}^{-1}$~\citep{2018MNRAS.478.3871W}.
The rate of period change should all be negative in this evolution model as
the star is contracting towards the sdOB branch with increasing stellar age.
If BLAPs are core helium-burning stars then the period of the radial
pulsation is related to the radius of the star, which is in turn
related to the helium abundance in the
core. Stellar evolution models computed
with \texttt{MESA} have been applied to OGLE-BLAP-011 assuming a core
helium-burning star revealing that the light curve can be reproduced
using first overtone radial oscillations~\citep{2019ApJS..243...10P}.

\citet[hereafter BJ20]{2020MNRAS.492..232B} extended the \texttt{MESA} models from BJ18
to include pre-WDs with masses from
$0.18\mathcal{M}_{\sun}$ to $0.46\mathcal{M}_{\sun}$ with the effects of radiative levitation.
They identified an extended region of iron-group instability using a non-adiabatic analysis of the \texttt{MESA} models at high time resolution (see Section 2 of BJ20 for details).
The observed periods of the BLAPs
from P17 and the high-gravity BLAPs are
consistent with the fundamental modes of \texttt{MESA} pre-WDs models.
The oscillation analysis by BJ20 indicate that pre-WDs with
effective temperatures of up to $\mathrm{T}_{\mathrm{eff}}\approx50,000$\,K can exhibit pulsations. Some of these models also show evidence of instability in higher-order p modes. Additional higher-order pulsation modes can be identified from high cadence light curves. If present, these multi-periodic pulsations would place further constraints on stellar evolution models from BJ20 and allow for analysis of the BLAP interior structure.

We structure the paper as follows. In \textsection2 we define the observations and data reduction on two OGLE-classified BLAPs during summer 2019 used in this analysis. In \textsection3 we present the method we used to analyse the light curves extracted from the data reduction with the goal of identifying additional periodicity in these variable stars. Finally, in \textsection4 we discuss the results of this analysis and constrain the potential pulsation modes present in the light curves of BLAPs. In this paper, the term \textit{amplitude} refers to the minimum to maximum variation in magnitude unless
otherwise stated.

\section{Follow-up Observations and Data Reduction}
The precise determination of the frequency spectrum of the
oscillations in the BLAPs can be used to identify the presence of
non-radial pulsations similar to those present in similar variable
stars on the H--R diagram. Frequencies which are different to the primary
radial frequency may be a result of harmonics of the primary mode, additional radial modes or non-radial modes. The harmonics of the primary mode have integer-multiple frequencies of the primary mode frequency as the periodogram fits the non-sinusoidal shape with additional sinusoidal components. Additional radial modes will feature a ratio with the primary frequency depending on the internal structure of the star. Any remaining frequencies are possibly non-radial modes with low amplitude, sinusoidal
shapes in the light curves. High cadence time-series can
be used in conjunction with signal-analysis techniques to reveal
the amplitude of any oscillations at a given frequency in the light
curves~\citep{2007CoAst.150..234K}. The frequencies which
can be exposed by such an analysis are limited by the cadence and
baseline of the light curves and care must be taken to avoid
introducing aliased frequencies into the analysis which can result
in ambiguity in the precise frequency of any detected
oscillations~\citep{2018ApJS..236...16V}. The uneven
sampling inherent to ground-based photometry is another concern
which has been addressed through the use of algorithms such as the
Lomb-Scargle~(LS) periodogram~\citep{1976Ap&SS..39..447L, 1982ApJ...263..835S}.
This method is equivalent to fitting
sinusoids to the light curve as a function of frequency whilst
incorporating a phase correction due to the uneven
sampling.

The light curves of the 14 candidate BLAPs discovered by OGLE
contain hundreds of I-band observations collected between 2001 and 2016. These
light curves are divided by the end of OGLE III in 2009~\citep{2008AcA....58...69U} and
the beginning of the OGLE IV in 2010~\citep{2015AcA....65....1U}. V-band observations are
also present in limited number which are used to determine
source colour and was sufficient for determining the change
in colour as a function of phase for the BLAPs suggesting the
likely radial pulsation source of their variability.

Data from P17 demonstrate the OGLE BLAP light curves have a long baseline but are based on low
cadence sampling with a median cadence of $2$
days. Whilst the uneven cadence of
these light curves does allow the detection of signals below the
nyquist sampling rate (such as the pulsation period which was
clearly detected for the discovery), it limits the reliability
of detections at shorter periods due to
aliasing~\citep{1949IEEEP..37...10S}. Using the RISE
instrument~\citep{2008SPIE.7014E..6JS} on the Liverpool
Telescope~\citep{2004SPIE.5489..679S}, we collected a high
cadence time-series of the two BLAPs OGLE-BLAP-009
and OGLE-BLAP-014.
Gaia DR2 data has determined that BLAP-009 is one of
the more luminous BLAP candidates~\citep{2018A&A...620L...9R}.
This research also indicates BLAP-014 is around the same colour,
but lower luminosity compared to BLAP-009. Line-blanketed
non-local thermodynamic equilibrium~(non-LTE) model atmospheres
computed on the spectroscopic followup from P17
suggest these two BLAPs have a similar surface gravity of
$\log g / ({\rm cm s^{-2}}) = 4.40\pm 0.18$ and $\log g / ({\rm cm s^{-2}}) = 4.42\pm 0.26$
respectively.

This new dataset consists of 3600 and 912 frames
were collected for BLAP-009 and BLAP-014. The observation strategy
is explained in the next section, to enable useful frequency analysis. The data was collected
during either dark or gray time with a seeing of $\approx1"$
under photometric conditions. See
Table~\ref{tab:obslog} for the details.The RISE instrument utilises
a single $720$\,nm long pass filter, corresponding to roughly
Sloan `i+z' filters. The multi-night data were
extracted with pyDIA~\citep{2009MNRAS.397.2099A, 2013MNRAS.428.2275B, 
2017zndo....268049M}, a difference image photometry package. This is
a software that can use Graphics Processing Units (GPUs) to efficiently
perform photometric extraction, in our case, an Nvidia GTX 1080 Ti. It
constructs optimal kernel models automatically for difference image
analysis that employs multiple kernel solutions and regularisation. This
method outperforms traditional photometric methods, particularly, in
crowded fields where the flux of multiple sources are extracted
simultaneously in order to arrive at accurate solutions. The light curves
are not absolute calibrated as they are only computed with respect to the
neighbouring stars in the CCD images. For this reason, the computed magnitude
values are not considered to be the true magnitudes, but the relative magnitudes
between the observations are valid for this analysis. The zero-point of the RISE
light curves used in this analysis are set by the default zero-point of pyDIA.

\begin{table*}
    \centering
    \begin{tabular}{c|cccccc}
         Target & Exposure Time & Night beginning on & No. of frames & Photometric\\\hline\hline
         \multirow{5}{*}{BLAP-009} & \multirow{5}{*}{10s} & 2019 Jun 12 & 720 & Y\\
                                   & & 2019 Jun 29 & 720 & Y\\ 
                                   & & 2019 Jul 01 & 720 & Y\\ 
                                   & & 2019 Jul 07 & 720 & Y\\ 
                                   & & 2019 Jul 09 & 720 & Y\\\hline
         \multirow{9}{*}{BLAP-014} & \multirow{9}{*}{30s} & 2019 Jul 26 & 120 & Y\\
                                   & & 2019 Jul 27 & 120 & Y\\ 
                                   & & 2019 Jul 28 & 120 & Y\\ 
                                   & & 2019 Jul 29 & 120 & Y\\ 
                                   & & 2019 Jul 31 & 120 & Y\\ 
                                   & & 2019 Aug 03 & 72 & Y\\
                                   & & 2019 Aug 04 & 120 & Y\\
                                   & & 2019 Aug 08 & 120 & Y\\
                                   & & 2019 Aug 09 & 120 & Y\\\hline
    \end{tabular}
    \caption{Observation properties for the frames used in the processing of the RISE BLAP-009 and BLAP-014 light curves.}
    \label{tab:obslog}
\end{table*}

The resulting single-band light curves were then saved into data
files for further analysis\footnote{The reduced
frames can be downloaded from the Liverpool Telescope archive at
the URL: \url{https://telescope.livjm.ac.uk/cgi-bin/lt_search}
under the proposal ID: \texttt{JL19A31}}. See attachment for the RISE photometry
of BLAP-009 and BLAP-014, and table A1 in Appendix A..

\section{Frequency Spectrum Analysis}
\subsection{Radial pulsation frequency}
Our first task was to independently identify the period of the primary
oscillation for the two BLAPs. To accomplish this we used the LS
periodogram to compute the frequency spectra. The LS periodogram is a relatively
light-weight method to compute a large frequency spectrum containing hundreds of
thousands of frequencies~\citep{1976Ap&SS..39..447L, 1982ApJ...263..835S}.
We used an R package implementation of the LS periodogram~\citep{1999ruf}\footnote{https://cran.r-project.org/web/packages/lomb/index.html}.
The output of this LS periodogram is a normalised LS power, a unitless quantity, computed by normalising against the variance of the input time-series.

The maximum period of the period search was set
to half of the baseline of each light curve where the baseline is defined as
$b = t_{\mathrm{max}}-t_{\mathrm{min}}$ where $t_{\mathrm{min}}$ is the
time instant of the initial observation and $t_{\mathrm{max}}$ is the
time instant of the final observation. This is the largest period which
can exhibit two complete cycles across the duration of the
observations, the minimum required by information theory.
The minimum period was defined by the Nyquist frequency of the multi-run
exposures. These periods are $20$\,s for BLAP-009 and $60$\,s for
BLAP-014. The steps between each of the frequencies in the frequency
spectrum were determined using equation~\ref{eq:ovfstep} with an
oversampling factor $v = 10$,
\begin{equation}
f_{\mathrm{step}}=\frac{1}{v(t_{\mathrm{max}}-t_{\mathrm{min}})}
\label{eq:ovfstep}
\end{equation}

\begin{figure}
 \includegraphics[width=\columnwidth]{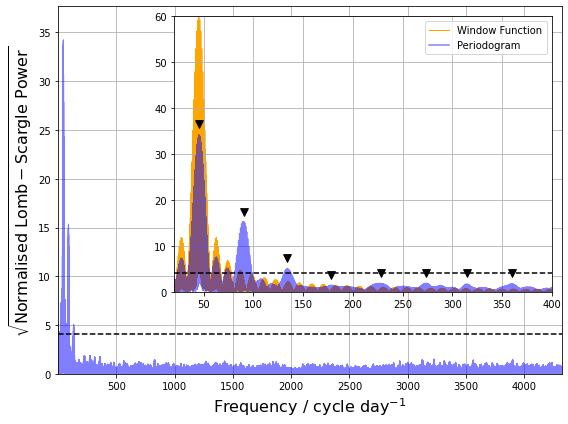}
 \caption{LS periodogram of the RISE light curve of BLAP-009. The
 frequency spectrum is dominated by the peak at a frequency of $45.091 \pm 0.0138$\,cycles $\mathrm{day}^{-1}$. This corresponds to a $31.935\pm 0.0098$\,mins primary
 pulsation period of the star. The inset focuses on the spectrum between $20-400$ cycles $\mathrm{day}^{-1}$ revealing the low frequency peaks. The spectral window function has been over-plotted and shifted to the primary frequency showing the associated sidelobes. Additional peaks are not aligned with the sidelobes and are highlighted by black markers. Their frequencies are shown to be harmonics of the primary frequency in table~\ref{tab:blapharm} and $2$ are significantly detected. The horizontal dashed line on both the main plot and the inset denote the $0.01$~($1\%$)
 significance level. All peaks above this line are considered to be
 significant.}
 \label{fig:blap009LS}
\end{figure}

\begin{figure}
 \includegraphics[width=\columnwidth]{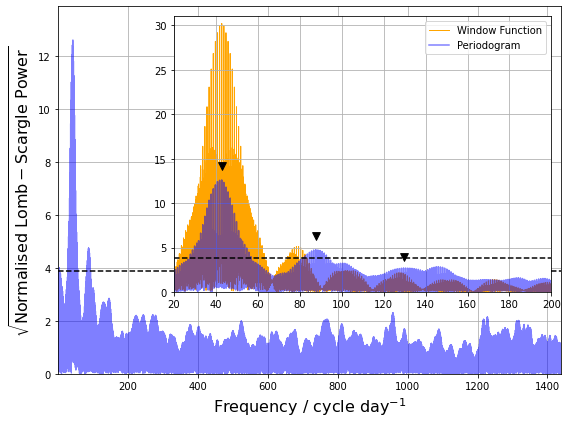}
 \caption{LS periodogram of the RISE light curve of BLAP-014. The
 frequency spectrum is dominated by the peak at a frequency of $42.826 \pm 0.0272$\,cycles $\mathrm{day}^{-1}$. This corresponds to a $33.625\pm 0.0214$\,mins primary
 pulsation period of the star. The inset focuses on the spectrum between $20-200$ cycles $\mathrm{day}^{-1}$ revealing the low frequency peaks. The spectral window function has been over-plotted and shifted to the primary frequency showing the associated sidelobes. There are less additional peaks than BLAP-009 but they also appear to be harmonics of the primary frequency although only $1$ is significant. They are highlighted by black markers and shown in table~\ref{tab:blapharm}. The horizontal dashed line on both the main plot and the inset denote the $0.01$~($1\%$)
 significance level. All peaks above this line are considered to be
 significant. The peaks are less significant due to the
 lower number of observations of BLAP-014. This figure has different scales relative to figure~\ref{fig:blap009LS}}
 \label{fig:blap014LS}
\end{figure}

Using these definitions for the minimum frequency, maximum frequency and the frequency steps, we computed a grid of candidate frequencies for the LS periodogram for the two RISE light curves time-series. The LS periodograms were then evaluated for BLAP-009 and BLAP-014 on their specific grid of candidate frequencies. This resulted in frequency spectra for BLAP-009 and BLAP-014 consisting of the candidate frequencies and their computed normalised LS power values. We refer to these frequency spectra as the BLAP periodograms from this point. The resulting periodograms are shown in figures~\ref{fig:blap009LS}
and \ref{fig:blap014LS} for BLAP-009 and BLAP-014 respectively. The
primary peaks are fine-tuned by a second LS periodogram with a frequency spectrum
focused around $\pm 1\%$ of the peak of the first LS periodogram with an
oversampling factor $v = 200$.
The resulting peaks agree with the period determined by OGLE III and IV
within the precision provided by our light curves. Our LS periodograms identify the period
of BLAP-009 as $31.935\pm 0.0098$\,mins and the period of BLAP-014 as
$33.625\pm 0.0214$\,mins. The uncertainty was defined as the half-width
half-maximum~(HWHM) of a Gaussian fit to the maximum peak although it
is important to note that this is not due to the precision of the
observations and mainly a function of the baseline of the light
curve~\citep{2018ApJS..236...16V}.


Dashed lines in figures~\ref{fig:blap009LS} and~\ref{fig:blap014LS} display the $0.01$~($1\%$) significance level determined
using a False Alarm Probability~(FAP). This quantity measures the
probability that, in the presence of no signal~(null hypothesis), a peak
of a given size may still result due to a coincidental alignment of
Gaussian distributed random errors~\citep{1982ApJ...263..835S}. Making
the assumption that the periodogram consists of a number of independent
frequencies, $N_{\mathrm{eff}}$, the significant LS power $\sigma$ is
calculated using equation~\ref{eq:sig},
\begin{equation}
\sigma = -\log{\left[1 - (1 - \alpha)^{\frac{v}{2 n}}\right]}
\label{eq:sig}
\end{equation}
where $\alpha = 0.01$ is the significance level, $n$ is the number of
frequencies in the frequency spectrum, determined by $f_{\mathrm{min}}$,
$f_{\mathrm{max}}$ and $f_{\mathrm{step}}$, and $v$ is the oversampling
factor.

We also computed the spectral window functions of the RISE light curves of BLAP-009 and BLAP-014. This calculation is shown in equation~\ref{eq:specwindow},
\begin{equation}
P_j = \frac{1}{N}\left|\sum\limits_{j=1}^N \exp\left({i 2\pi f_j t}\right)\right|^2
\label{eq:specwindow}
\end{equation}
where $P_j$ is the spectral power of the $j^{\mathrm{th}}$ candidate frequency, $N$ is the number of observations in the light curve, $i$ is the imaginary unit, $f_j$ is the $j^{\mathrm{th}}$ candidate frequency in the frequency spectrum and $t$ are the time instants of the light curve observations.

This computation is offset to the primary frequency of the two periodograms and over-plotted on the inset plots of figures~\ref{fig:blap009LS} and \ref{fig:blap014LS}. The window function reveals that many of the smaller low frequency peaks for both periodograms are a result of interference from sidelobes in the window function due to the sampling times of the RISE light curves. The signals at these frequencies are aliased from the signal at the primary frequency.

In addition to over-plotting the spectral window function, we also investigated the aliased peaks around the pulsation period from the fine-tuned LS periodograms. These peaks are close to the primary frequency as they are aliased by low frequency sampling periodicity. For BLAP-009, the largest of the peaks surrounding the primary frequency
correspond to an alias
with a $0.997$\,day period ($1.003$ cycles $\mathrm{day}^{-1}$ frequency), the sidereal day. Other peaks are associated
with $0.5$ and $2$ multiples of the sidereal day. The final spurious
period causing detectable aliased peaks is approximately $8.2$\,days ($0.122$ cycles $\mathrm{day}^{-1}$ frequency). An
investigation of this period indicates that it is a result of the schedule of the observation nights. BLAP-014 also shows aliased peaks due to the
sidereal day but of weaker power relative to the main pulsation period
peak compared with BLAP-009. There also did not appear to be any
significant aliasing from a longer sampling period for this light curve.

The periodograms of both BLAP-009 and BLAP-014 do not display any significant high frequency peaks. This indicates the lack of detection of an additional periodic signal. We investigate the limits of this non-detection in the next section. There are a number of additional peaks at lower frequencies once the interference lobes from the spectral window function were identified. Table~\ref{tab:blapharm} show these frequencies for BLAP-009 and BLAP-014. They are also highlighted in the inset plots of figures~\ref{fig:blap009LS} and \ref{fig:blap014LS} by black markers. These peaks are at or near integer multiples of the primary frequency and are associated with higher order harmonics of the primary frequency. This is a result of non-sinusoidal sawtooth signals requiring
higher order harmonics to fit the characteristic shape with a set of sinusoids. BLAP-009 displays $3$ harmonic frequencies with significant peaks and for BLAP-014 there are $2$ significant harmonic frequencies. Determining the significance of these harmonic frequencies is important as it determines how many harmonics must be modelled to fit the light curve. The $4^{\mathrm{th}}$ harmonic frequency of BLAP-009 is interesting as it has a smaller peak in the periodogram than the $5^{\mathrm{th}}$ to $8^{\mathrm{th}}$ harmonic frequencies. This is an indication that the BLAP-009 light curve may contains features which are better modelled by these higher harmonics although their peaks are not significant. The $3^{\mathrm{rd}}$ harmonic frequency of BLAP-014 also coincides with a sidelobe in the spectral window function which reduces the confidence in this detection.
\begin{table}
    \centering
    \begin{tabular}{cccc}
        & & \\\hline\hline
        & OGLE-BLAP-009 & \\\hline\hline
        ID & Frequency (cycles $\mathrm{day}^{-1}$) & Ratio to $F_1$ & Significant\\\hline
        $F_1$ & $45.091$ & $1.000$ & \checkmark\\
        $F_2$ & $90.181$ & $2.000$ & \checkmark\\
        $F_3$ & $134.276$ & $2.978$ & \checkmark\\
        $F_4$ & $178.360$ & $3.956$ & $\times$\\
        $F_5$ & $227.950$ & $5.056$ & $\times$\\
        $F_6$ & $273.554$ & $6.067$ & $\times$\\
        $F_7$ & $314.634$ & $6.978$ & $\times$\\
        $F_8$ & $359.724$ & $7.978$ & $\times$\\\hline\hline
        & OGLE-BLAP-014 & \\\hline\hline
        ID & Frequency (cycles $\mathrm{day}^{-1}$) & Ratio to $F_1$ & Significant\\\hline
        $F_1$ & $42.826$ & $1.000$ & \checkmark\\
        $F_2$ & $87.660$ & $2.047$ & \checkmark\\
        $F_3$ & $129.492$ & $3.024$ & $\times$\\
    \end{tabular}
    \caption{Frequencies of harmonic peaks in the periodograms of BLAP-009 and BLAP-014 highlighted by black markers in the inset plots of figures~\ref{fig:blap009LS} and \ref{fig:blap014LS}. The first $3$ harmonic frequencies of BLAP-009 and the first $2$ harmonic frequencies of BLAP-014 have corresponding periodogram peaks stronger than the $0.01$~($1\%$) significance level.}
    \label{tab:blapharm}
\end{table}

The light curves for the two BLAPs were then epoch-folded using equation~\ref{eq:epochfold} to reveal the shape of the primary oscillation,
\begin{equation}
\phi_i(P)=\mathrm{mod}\left[\frac{t_i - t_0}{P}\right]
\label{eq:epochfold}
\end{equation}
where $\phi_i(P)$ is the phase value of the observation $i$ as a
function of the candidate period $P$, $t_i$ is the measurement time of
the observation $i$, $t_0$ is an arbitrarily chosen start
time~(currently defined such as at the Heliocentric Julian Date (HJD) of $0.0$ corresponds to a
phase of 0.0), $P$ is a candidate period for the light curve and the
modulus~(mod) operation retains the decimal component of the
calculation~\citep{1996A&AS..117..197L}. 

\begin{figure}
 \includegraphics[width=\columnwidth]{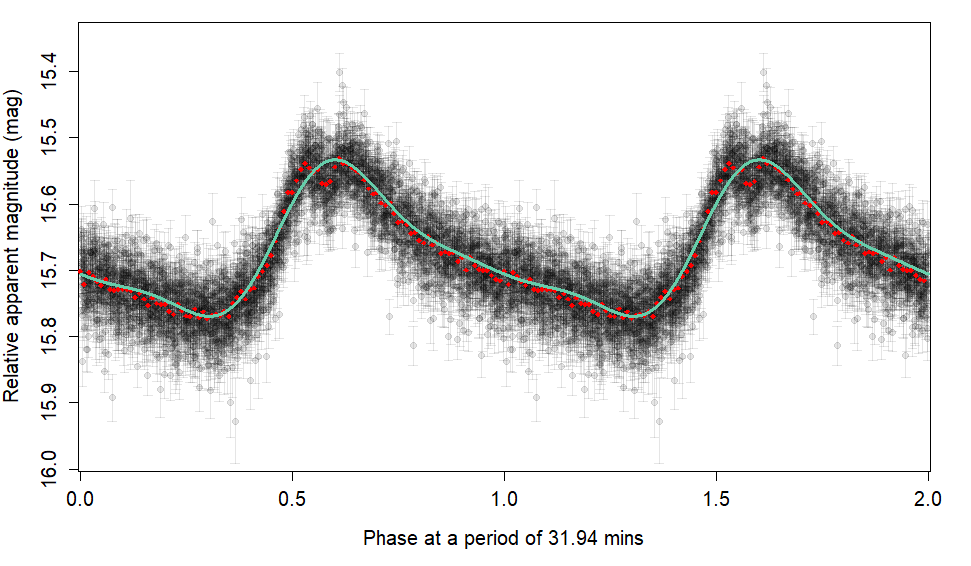}
 \caption{Scatter plot in black shows the folded light curve of BLAP-009
 at a period of $31.935$\,mins. Weighted average of the light curve with $100$ phase bins
 is shown in red and a fit from a $3$-harmonic Fourier model (see table~\ref{tab:coeffs}) is shown in light blue (see online version for colour).}
 \label{fig:blap009LC}
\end{figure}

\begin{figure}
 \includegraphics[width=\columnwidth]{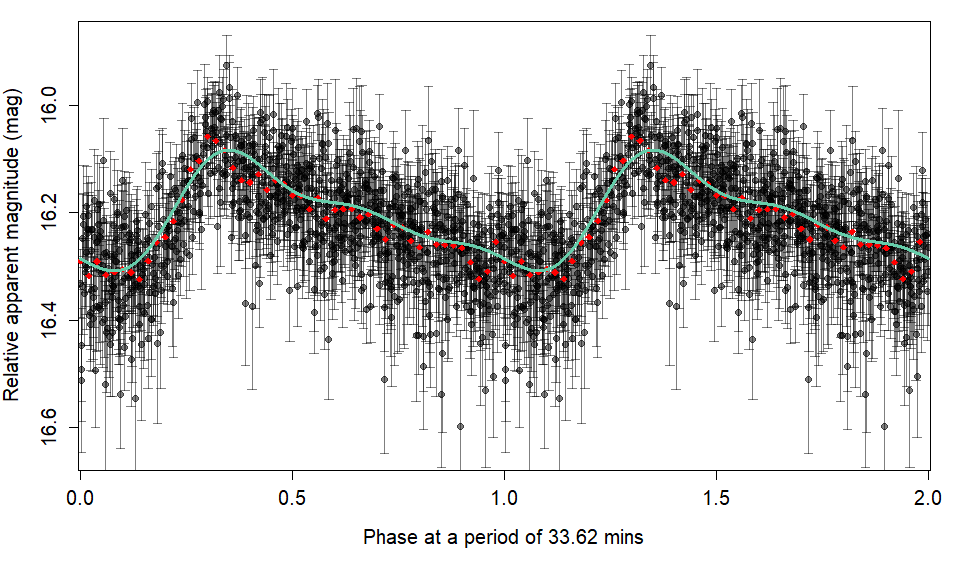}
 \caption{Scatter plot in black shows the folded light curve of BLAP-014
 at a period of $33.625$\,mins. Weighted average of the light curve with $50$ phase bins
 is shown in red and a fit from a $3$-harmonic Fourier model (see table~\ref{tab:coeffs}) is shown in light blue (see online version for colour).}
 \label{fig:blap014LC}
\end{figure}

The epoch-folded light curves of BLAP-009 (shown in figure~\ref{fig:blap009LC}) and BLAP-014 (shown in figure~\ref{fig:blap014LC}) reveal a characteristic sawtooth shape further reinforcing the conclusion that the additional periodogram peaks are harmonics. The epoch-folded light curves were also phase binned into 100 bins using a weighted mean to denoise the light curve further highlighting the oscillation. These binned data points are shown as red diamonds in the figures and the light blue lines are the result of $3$-harmonic Fourier fits to the two light curves described in the next section (see online version for colour). BLAP-009 exhibits an interesting and unique feature at maximum light where there appear to be two maxima. This may be a result of one peak corresponding to maximum temperature, and the other for maximum radius. This feature also appears to be present in the OGLE light curves of BLAP-009 but, as of current data, appears to be unique to this BLAP. This may also be the cause for the stronger higher-order harmonic frequency peaks in the BLAP-009 periodogram.

\subsection{Search for other signals}
To identify further periodic signals in the data, the dominant period
in the frequency spectrum must be subtracted out from the light curve in
a process named \textit{prewhitening}. This is accomplished by fitting a periodic
model with the dominant period as the argument. In the case of a
sinusoidal signal, a simple sinusoidal model with an amplitude and phase
is all that is required. The signals from BLAP-009 and BLAP-014 are
clearly non-sinusoidal and appear sawtooth in shape as exhibited by the harmonic peaks in the periodograms; therefore, we select a multi-harmonic Fourier model
in order to fit this shape~\citep{2009A&A...507.1729D, 2011ApJ...733...10R, 2012ApJS..203...32R}. The model is shown in
Equation~\ref{eq:fouriermodel} with $n = 3$ as supported 
by the number of significant harmonic peaks in the periodogram of BLAP-009,
\begin{equation}
m_i=\sum\limits_{j=1}^n \big[a_j\sin(2\pi jf t_i)+b_j\cos(2 \pi jf t_i)\big]+b_0
\label{eq:fouriermodel}
\end{equation}
where $b_0$ is the mean magnitude of the light curve, $m_i$ is the model magnitude of time
instant $t_i$ where $i$ identifies the $i^{\mathrm{th}}$ data point,
$a_j$ and $b_j$ are Fourier coefficients for the fitted model and
$f$ is the dominant frequency, $f = P^{-1}$ where $P$ is the previously
identified period. This model has $7$ coefficients which includes the intercept
with an additional $3$ sine and $3$ cosine components
which model the amplitudes and phases of the three sinusoidal harmonics.

Using higher-order harmonics carries an additional risk due to potential
over-fitting on noise. To mitigate this, we utilise a
regularised least-squares fitting technique to apply a weighting to each harmonic proportional
to $j^4$ where $j$ is the harmonic number of a given sinusoid. Equation~\ref{eq:reg}
is the function to be minimised to find the optimal model,
\begin{equation}
R\left(\theta, \lambda\right) =\sum\limits_{i=1}^N \frac{\left(d_i-\left(\theta^{\top}t_i\right)\right)^2}{\sigma_i^2} + N\lambda \sum\limits_{j=1}^n j^4\left(a_j^2 + b_j^2\right)
\label{eq:reg}
\end{equation}
where $\theta$ is a vector of the model parameters, $\lambda$ is the regularisation
parameter, $N$ is the number of light curve points, $d_i$ are the
photometric magnitude data points, $t_i$ are the photometric time instants and
$\sqrt{a_j^2 + b_j^2}$ is the amplitude of the $j^{\mathrm{th}}$ Fourier
harmonic and $n=3$ for the $3$-harmonic model. The value of the regularisation parameter allows the control
of the smoothing of the model with small values allowing the modelling
of high frequency structure and large values smooth this structure out.
For these light curves the regularised fit is applied with a
regularisation parameter of $0.01$. Table~\ref{tab:coeffs} shows the 
coefficients of the models fit to the BLAP-009 and BLAP-014 light curves.
\begin{table}
    \centering
    \begin{tabular}{ccc}
        Coefficient & OGLE-BLAP-009 & OGLE-BLAP-014\\\hline\hline
        $f$ (cycles/day)    & $45.0909383371$ & $42.8257928504$\\
        $c$ (mag)    & $15.669736901$ & $16.205144758$\\\hline
        $a_1$ (mag)    & $8.3392464\times10^{-2}$ & $-2.7476492\times10^{-2}$\\
        $b_1$ (mag)    & $4.7774440\times10^{-2}$ & $8.1055037\times10^{-2}$\\\hline
        $a_2$ (mag)    & $-3.7053363\times10^{-2}$ & $3.4618251\times10^{-2}$\\
        $b_2$ (mag)    & $-1.7831991\times10^{-2}$ & $1.7012638\times10^{-2}$\\\hline
        $a_3$ (mag)    & $1.0565454\times10^{-2}$ & $7.773188\times10^{-3}$\\
        $b_3$ (mag)    & $6.159469\times10^{-3}$ & $-1.7585455\times10^{-2}$\\\hline
    \end{tabular}
    \caption{Coefficients of regularised $3$-harmonic Fourier models of BLAP-009 and BLAP-014,
    with periods of $31.935$ and $33.625$\,min respectively. The meanings of these coefficients
    are shown in equation~\ref{eq:fouriermodel}.}
    \label{tab:coeffs}
\end{table}

This model also allows for the computation of the amplitude of the primary radial
amplitude reducing contamination from noise in the observations. Using the coefficients in
table~\ref{tab:coeffs} the amplitude of the BLAP-009 pulsation in the RISE $720$\,nm filter
is $0.236$\,mag. A similar computation using the model of BLAP-014 provides an
amplitude of $0.225$\,mag. These amplitudes are below those reported by P17
from the OGLE I-band light curves which may be a result of noise or the longer
wavelengths probed by the RISE $720$\,nm filter. We also find that our light curves show the
oscillations in this band of BLAP-009 are higher amplitude than BLAP-014, disagreeing with the values
reported by P17. This discrepancy is likely a result of the difference in the
two methods used to determine these values or the differences in the filter bands of OGLE and RISE.

\begin{figure}
 \includegraphics[width=\columnwidth]{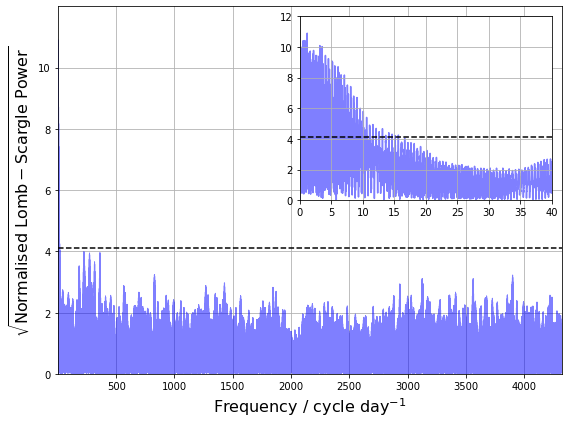}
 \caption{LS periodogram of the residual light curve of BLAP-009 after
 prewhitening at a period of $31.935$\,mins. The frequency spectrum
 is dominated by a set of low frequency peaks shown in the inset plot.
 These peaks were
 examined and found to be a result of a sampling frequency causing
 spurious alignment of the data points. The set of peaks near the significance level between $0-500$\,cycles $\mathrm{day}^{-1}$
 are higher-order harmonics which were not prewhitened by the $3$-harmonic model.
 The dashed line denotes the $0.01$~($1\%$) significance level.}
 \label{fig:blap009LSres}
\end{figure}

\begin{figure}
 \includegraphics[width=\columnwidth]{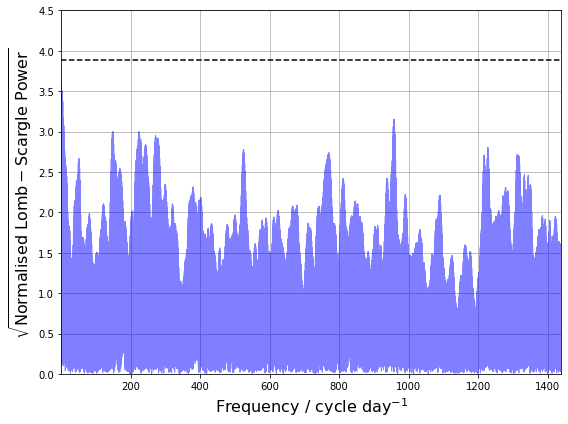}
 \caption{LS periodogram of the residual light curve of BLAP-014 after
 prewhitening at a period of $33.625$\,mins. The larger low frequency
 peak is again associated with a spurious sampling frequency. None of
 the frequencies in the residual spectrum have a significant detection.
 The dashed line denotes the $0.01$~($1\%$) significance level.}
 \label{fig:blap014LSres}
\end{figure}

The fitted $3$-harmonic Fourier model is then subtracted from the
original light curve to leave the prewhitened, residual light curve. We
then recomputed the LS periodogram on these residuals to search for
additional periodic signals shown in figures~\ref{fig:blap009LSres} and
\ref{fig:blap014LSres}. Whilst at first glance, it does appear that
BLAP-009 has a significant low frequency peaks, these were
determined to be a result of the sampling cadence by epoch-folding at
these frequencies revealing extremely poor phase coverage of the
resulting folded light curve. Periods over $2$\,hrs were highly
affected by these sampling artifacts. This is an interesting result as
each individual night had a $2$\,hr duration, therefore, this duration
is the longest period with a guaranteed complete phase coverage.
BLAP-014 also shows a low frequency correlated noise component due to
sampling, but it is much less pronounced. The observations for
BLAP-014 were over a more regular cadence than BLAP-009 which results
in a cleaner power spectrum despite the shorter $1$\,hr duration of
nightly observations. After the explanation of the significant peaks
due to correlated noise, the remaining peaks are substantially below
the $0.01$~($1\%$) significance level. A few of the larger peaks had their
associated frequencies epoch-folded with the prewhitened light curves
but no unambiguous signal was found by inspection. Fits of the $3$-harmonic
Fourier model at these frequencies on the prewhitened data reveal
amplitudes of around $10$\,mmag for BLAP-009 and $40$\,mmag for
BLAP-014 but, despite the regularisation, appear to be fitting noise.

\subsection{Null Hypothesis with Artificial Signal Injection}
In order to characterise the detection limits of our data as both
functions of amplitude and frequency, we inject artificial sinusoidal
signal to the data and explore the parameter space in which the signal
can be identified with the methodology outlined in \textsection3. A
grid of $9\times11$ period-amplitude pairs were used for the hypothesis
tests of null signals. Any undetected variability is likely to be low
amplitude which is usually due to non-radial oscillations which are
normally sinusoidal in shape. Due to this, the signals injected into the
data are purely sinusoidal. The LS periodogram may still detect a non-sinusoidal signal but at a lower confidence.

The periods selected for the artificial signals are [$2$, $5$, $10$,
$20$, $30$, $60$, $120$, $300$ and $600$]\,mins. The amplitudes of the
artificial sinusoids are [$1$, $2$, $5$, $10$, $15$, $20$, $25$, $30$,
$40$, $50$ and $100$]\,mmag. For each combination of period and
amplitude, an artificial sinusoidal signal is injected into the light
curve and a LS periodogram is computed on the data with a minimum
frequency $f_{\mathrm{min}}$ equal to half of the baseline of the light curve
and a maximum frequency $f_{\mathrm{max}} = 900$, large enough to
include the $2$\,min periodic signal with an oversampling factor
$v = 10$. The computed period was fit with a $3$-harmonic Fourier
model and the light curve prewhitened using this model as described
in \textsection3. A second LS periodogram is computed on the residual
light curve and the LS power of the candidate frequency closest to the period of the
artificial signal was calculated along with the significance level.

\begin{figure}
 \includegraphics[width=\columnwidth]{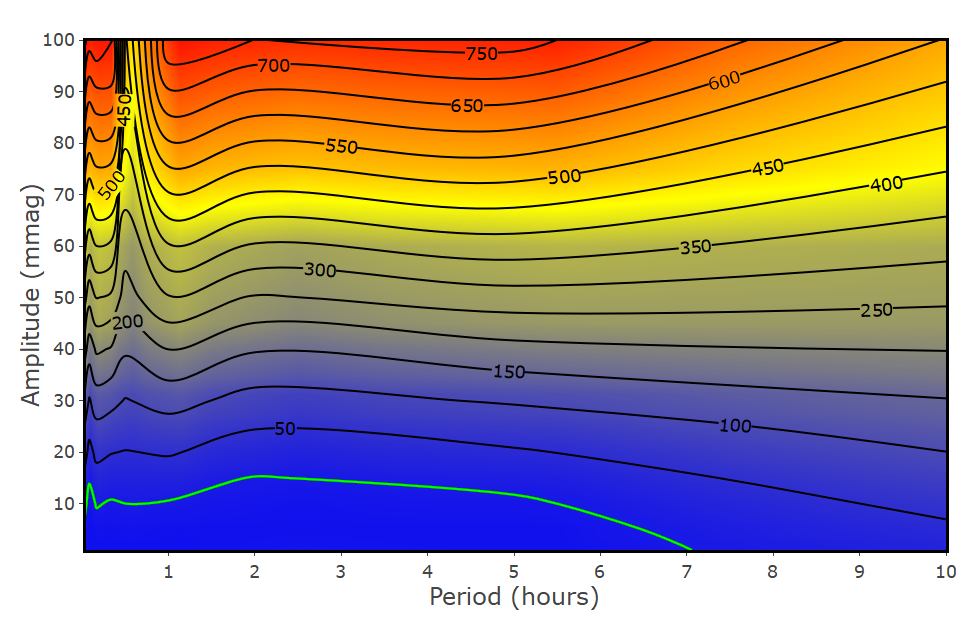}
 \caption{Contour plot of the normalised LS power interpolated over the $9$-amplitude
 by $11$-period grid of sinusoidal signals injected into the RISE light curve of BLAP-009.
 The green line (see online version for colour) is the contour line denoting
 the $0.01$~($1\%$) significance level, above which, in the y-axis, the
 injected signal can be detected.}
 \label{fig:blap009_det}
\end{figure}

\begin{figure}
 \includegraphics[width=\columnwidth]{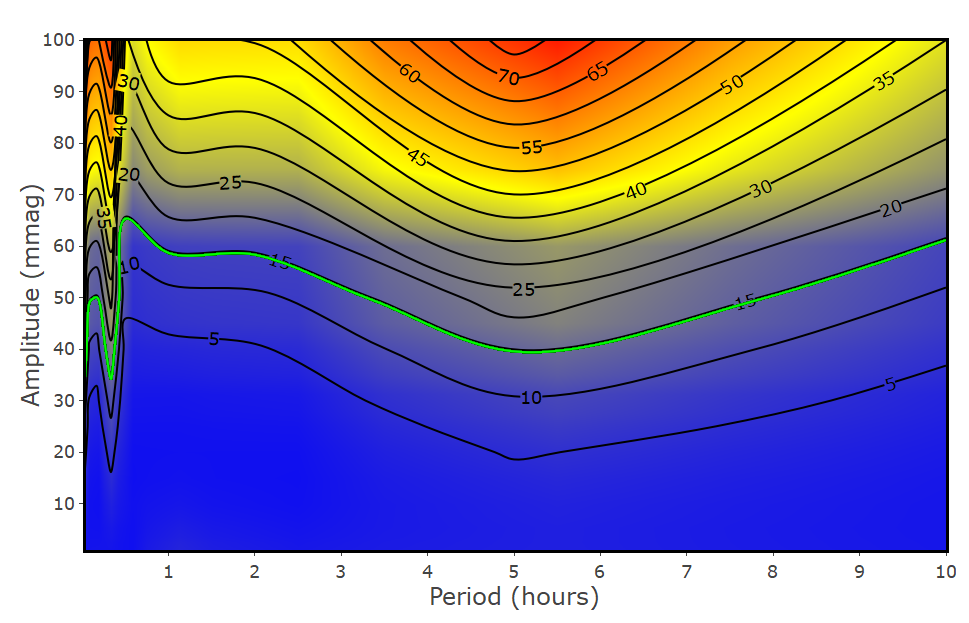}
 \caption{Contour plot similar to Figure \ref{fig:blap009_det} for BLAP-014. The green line
 (see online version for colour) is the contour line denoting
 the $0.01$~($1\%$) significance level, above which, in the y-axis, the
 injected signal can be detected.}
 \label{fig:blap014_det}
\end{figure}

The contour plots shown in Figures~\ref{fig:blap009_det} and
\ref{fig:blap014_det} show the results of this analysis for BLAP-009
and BLAP-014 respectively. The contours indicate the interpolated LS
power across the period-amplitude grid. The light-green line~(see
online version for colour) indicates the contour of the $0.01$~($1\%$) significance level
computed using equation~\ref{eq:sig} and is analogous to the dashed lines
for the confidence limit in figures~\ref{fig:blap009LS} and \ref{fig:blap014LS}.
The contours are interpolated
across the period-amplitude grid using the Marching Squares algorithm.

\begin{figure}
 \includegraphics[width=\columnwidth]{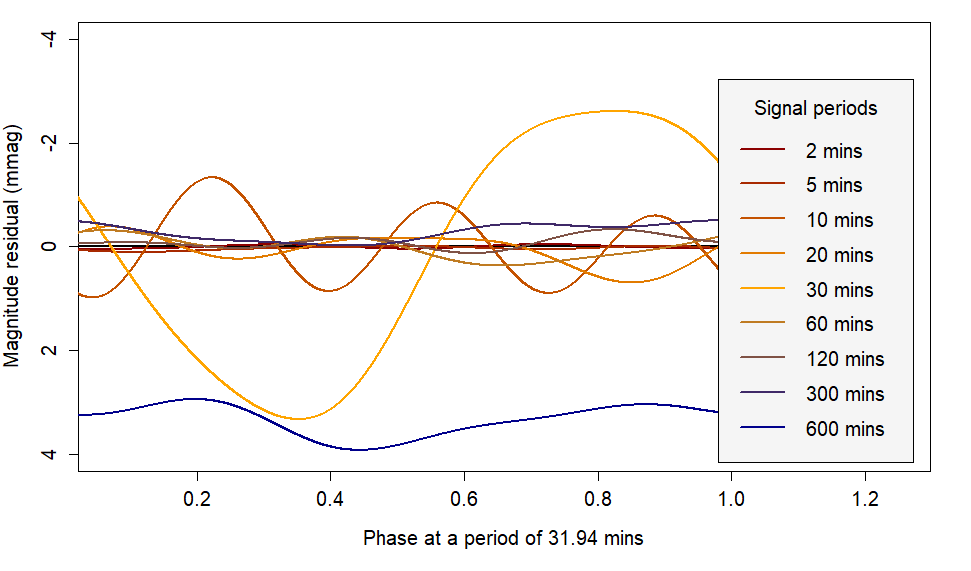}
 \caption{Plot of the magnitude residuals of the $3$-harmonic Fourier
 fit to the BLAP-009 light curve at the pulsation period with an
 injected $15.20$\,mmag signal at $9$ different periods relative to the
 observed BLAP-009 light curve. The plot demonstrates that the injected
 signal can perturb the model fit to the pulsation resulting in
 unexpected features in Figures~\ref{fig:blap009_det} and
 \ref{fig:blap014_det}. The largest perturbation occurs at periods
 close to the pulsation period where the amplitude of the residual is
 $5.94$\,mmag for a $15.20$\,mmag signal, a substantial fraction of the
 injected signal.}
 \label{fig:blap009co}
\end{figure}

The results for BLAP-009 show that any periodic sinusoidal signal can
be detected down to a lower threshold than BLAP-014 which is not
surprising as BLAP-009 was observed over a longer baseline and with
many more individual observations with smaller uncertainties. The strength of the peaks of the
pulsation periods was sufficient to not be affected by the injected
artificial signal. Therefore, the periods determined by the first LS
periodograms were independent of the period and amplitude of the
artificial signals. The same cannot be said for the quality of the
$3$-harmonic Fourier models fit to the pulsation periods.
Figure~\ref{fig:blap009co} reveals the magnitude residuals of the fit
to the BLAP-009 pulsation period with injected $15.20$\,mmag
sinusoidal signals (the lowest amplitude signal which produces a significant $0.01$~($1\%$)
peak in the LS periodogram) for the nine different periods from the grid in
Figure~\ref{fig:blap009_det}. Despite fitting the same period, the
coefficients of the Fourier models differ slightly due to the effect
of the injected signal which results in the residual signal.

The model with an injected signal of period $2$\,min is almost identical
to the initial model and shows the parameters of the Fourier model when
unperturbed by the injected signal. The $30$\,min injected signal
residuals show this model is clearly perturbed by the presence of the
injected signal as it's period is similar to the $31.935$\,min period of
the pulsation. The $5.94$\,mmag amplitude of this residual is a
substantial fraction of the injected $15.20$\,mmag signal. As a result,
the Fourier model is under-fitting the pulsation signal and removing
some of the injected signal. We conclude that this is the mechanism
responsible for the contour feature in Figures~\ref{fig:blap009_det}
and \ref{fig:blap014_det} overestimating the LS power of periods close
to the pulsation period of the BLAPs. The $600$\,min injected
period light curve model shows a clear variation in the mean magnitude
of it's Fourier fit. This appears to be caused by the sampling of the
observations of BLAP-009 as the phase coverage of the light curve decreases above periods of $2$\,hrs
causing inhomogeneous groups of data points when the light curve is
phased at the pulsation period. It is likely that this complication to
the Fourier fit, in combination with the poor phase coverage at longer
periods producing a correlated noise, results in the unexpected contours
in figure~\ref{fig:blap009_det} at longer periods where the significance
contour decreases in amplitude until it contacts the zero signal margin.
This result suggests that a non-existent signal is significantly
detected which is a nonsensical notion. An inspection of the same
feature in the BLAP-014 Fourier fit coefficients reveals this effect is
substantially less pronounced for the BLAP-014 light curve reinforced
by the lack of contour decay at long periods in
Figure~\ref{fig:blap014_det}. These features arise as the significance
contour is constructed using the FAP which assumes a purely Gaussian
noise which does not hold for these light curves.

Taking into account these perturbations, any sinusoidal signal above
$15.20\pm0.26$\,mmag is significantly detectable to the $0.01$~($1\%$) significance
level in the RISE BLAP-009 light curve. Similarly, any sinusoidal signal
above $58.60\pm3.44$\,mmag is significantly detectable to the $0.01$~($1\%$)
significance level in the RISE BLAP-014 light curve. These boundaries
can vary as a function of the period of both the pulsation and the
secondary lower-amplitude signal. We can confidently state that our light
curves do not show evidence of any additional periodicity given the above
confidence limits down to $20$\,s for BLAP-009 and $60$\,s for BLAP-014.
Additionally, we did not detect any non-significant peak which, when
epoch-folded, suggested any additional variability. The only detected
signals were strongly significant peaks at the two previously identified
periods from the OGLE survey.

\section{Conclusion and Discussion}
This work reports a null multi-periodicity of BLAP-009 down to a
$20$\,s period and BLAP-014 down to a $60$\,s period with no additional
modes with amplitudes greater than $15.20\pm0.26$\,mmag for BLAP-009
and $58.60\pm3.44$\,mmag for BLAP-014. The upper limits of the detectability
are heavily influenced by the correlated noise in the light curves produced by
the sampling cadence. For BLAP-009 any signal with a period $\gtrapprox7$\,hrs has a
peak greater than the $0.01$~($1\%$) significance level. This can be clearly seen
in figures~\ref{fig:blap009LSres} and \ref{fig:blap009_det}. Meanwhile,
BLAP-014, whilst having a greater limiting amplitude due to less
observations has less correlated noise. As a result, we can
conclude there are no additional signals up to half of the baseline of the light curve at $6.48$\,days. Independently, an investigation of the aliased features around the dominant pulsation
period and the computation of spectral window functions reveal they are all associated with known spurious periods. This work can, however,
provide a strong prior in future work in determining the pulsation mode
when coupled with multi-band photometry or time-resolved medium/high
resolution spectra.

Our data indicate that it is unlikely for there to be additional modes
in BLAP-009 and BLAP-014. Radial oscillations in more than one of the fundamental, first or second overtones simultaneously should be detectable given
the computed detection limits of the BLAP-009 light curve.
Stellar theory shows that multi-mode pulsators have well-defined period ratios between their
fundamental and overtone modes such as $P_1/P_0 = 0.71$ for double-mode Cepheids in the Milky Way~\citep{1966VA......8..191P}
and $P_1/P_0 = 0.746$ for double-mode RR Lyraes in globular clusters~\citep{1993Ap&SS.210.....T}.
The structure of BLAPs has not been confidently determined given the likely presence of
a degenerate helium core. Despite this, if we assume a similar ratio 
to the double-mode Cepheids for BLAP-009 and that the
detected period is a fundamental mode, the expected period $P_1$ of the first overtone would be
$22.674$\,mins or a frequency of $63.508$\,cycles $\mathrm{day}^{-1}$. This frequency should be
detectable with our current light curves and is not interfered with by the harmonics of the primary
signal. Alternatively, if the detected period is the first-overtone, the fundamental mode period is
$44.980$\,mins or a frequency of $32.015$\,cycles $\mathrm{day}^{-1}$. In this case, the harmonics of
the primary signal are also not interfering with the detection of this signal although the window
function is much stronger at this frequency for the RISE BLAP-009 light curve.
Extrapolating this as common to all BLAPs, the
oscillation appears to be a single radial p mode of either the
fundamental mode, or the first overtone~\citep{2018MNRAS.481.3810B, 2019ApJS..243...10P}.
Other important quantities to refine are the current period of these two BLAPs and the
rate of period change relative to the previous OGLE surveys. These values can be used to refine the
evolutionary status of these stars as the period of radial pulsation modes is dependent on the mean stellar density~\citep{1917Obs....40..290E}.

Our precision of the period estimation is lower than the OGLE surveys due to the shorter baseline
of the observations as this is the property of the light curve which defines the
uncertainty in the frequency spectrum via the Rayleigh frequency resolution: $\delta f = \frac{1}{T}$, where $\delta f$ is the uncertainty in a frequency measurement and $T$ is the baseline of the light curve.
The larger uncertainties on the periods from the frequency spectra than the OGLE III and IV surveys make it difficult to make conclusions about the rate of period change from our RISE light curves.
The faint magnitude of the known candidate BLAPs, even the brightest target
BLAP-009, results in substantial difficulty to produce both high
cadence light curves with a 2-meter telescope pointing at high air mass whilst maintaining the required signal-to-noise to probe low amplitude oscillations of the order of
hundredths of a magnitude.

Future photometric studies on BLAPs would be greatly enhanced through
the identification of additional candidates. Given the null detection
of BLAPs in LMC and SMC fields from the OGLE
surveys~\citep{2018pas6.conf..258P}, the Omega-White
survey~\citep{2015MNRAS.454..507M} and the ZTF high-cadence survey at
low Galactic latitudes with ZTF~\citep{2019PASP..131a8002B,
2019PASP..131f8003B, 2019PASP..131g8001G} are
the most likely ongoing surveys that will yield new BLAPs.

\section*{Acknowledgements}
The Liverpool Telescope is operated on the island of La Palma by
Liverpool John Moores University in the Spanish Observatorio del
Roque de los Muchachos of the Instituto de Astrof{\'i}sica de
Canarias with financial support from the UK Science and Technology
Facilities Council.

PRW acknowledges financial support from the Science and Technology Facilities Council (STFC).

ML acknowledges financial support from the OPTICON.

PRW thanks the useful discussions with Gavin Ramsay, 
Simon Jeffrey and Conor Byrne from the Armagh Observatory and the recommendations of Alejandra Romero from the Universidade Federal do Rio Grande do Sul.

PRW thanks the comments of the anonymous reviewer contributing to the improvement of this manuscript.  








\appendix
\section{RISE Photometry}
\label{sec:relphot}
Table~\ref{tab:relphot} describes the content of the supplementary
material available online for BLAP-009 and BLAP-014.
\begin{table}
    \centering
    \begin{tabular}{c|l}
        Column & Description\\\hline
        $1$    & Heliocentric Julian Day~(HJD)\\
        $2$    & Phase\\
        $3$    & Relative Photometry / mag\\
        $4$    & Uncertainties / mag \\\hline
    \end{tabular}
    \caption{The photometry of BLAP-009 and BLAP-014,
    with periods of $31.935$ and $33.625$\,min respectively.
    Phase 0.0 is assigned arbitrarily and is defined as phase = $0.0$ at HJD = $0.0$.}
    \label{tab:relphot}
\end{table}

\bsp	
\label{lastpage}
\end{document}